\documentclass{cimento}
\usepackage{graphicx}

\title{Principal Component Analysis of Gamma-Ray Bursts' Spectra}
\author{
Z.~Bagoly\from{ins:elte}\ETC,
I.~Horv\'ath\from{ins:bolyai},
L.~G.~Bal\'azs\from{ins:konkoly},
L.~Borgonovo\from{ins:stockholm},
S.~Larsson\from{ins:stockholm},
A.~M\'esz\'aros\from{ins:prague}
        \atque
F.~Ryde\from{ins:stockholm}
}

\instlist{ 
\inst{ins:elte} Laboratory for Information Technology, E\"{o}tv\"{o}s
University, H-1117 Budapest, P\'azm\'any P. s.  1./A, Hungary.
\inst{ins:bolyai} Dept. of Physics, Bolyai Military Univ. H-1456 Budapest, POB 12, Hungary
\inst{ins:konkoly} Konkoly Observatory, H-1525 Budapest, POB 67, Hungary
\inst{ins:stockholm} Stockholm Observatory, AlbaNova, SE-106 91 Stockholm,
Sweden
\inst{ins:prague} Astronomical Institute of the Charles University, V
Hole\v{s}ovi\v{c}k\'ach 2, CZ-180 00 Prague 8, Czech Republic
}
\PACSes{\PACSit {98.70.-f}{98.70.Rz}{98.80}}
\begin{document}

\maketitle

\begin{abstract}
Principal component analysis is a statistical method, which
lowers the number of important variables in a data set. The use of this 
method for the bursts' spectra and afterglows is discussed in this paper.
The analysis indicates that three principal components are 
enough among the eight ones to describe the variablity of the data. 
The correlation between spectral index $\alpha$ and the redshift 
suggests that the thermal emission component becomes more dominant at 
larger redshifts.
\end{abstract}

\section{Introduction}

The principal component analysis
(PCA) is a powerful statistical method, if the observational data 
contain several variables \cite{ref:jollife}. If 
there are some correlations among the variables, then PCA lowers 
the number of variables needed for the description of the dataset without 
any essential loss of generality. It introduces the quantities
called principal components (PCs). Then one can omit 
some PCs not leading to essential loss of information about the dataset
(for more details about PCA see \cite{ref:jollife}).

In the topic of gamma-ray bursts (GRBs) the paper
 \cite{ref:bagoly98} used PCA on the dataset of 
BATSE Catalog ~\cite{ref:mee}. This Catalog contains 
nine measured variables in the gamma-band alone (three different peak fluxes, 
the four fluences in the four different energy channels, and the two 
durations) for any GRB. \cite{ref:bagoly98} has shown that
from these nine variables only three ones are also enough 
(one of the durations, a combination of the fluences and the peak fluxes, 
and the fluence on the highest energy channel alone).

As far as it is known, since 1998 no further PCA was applied 
on any GRBs' datasets. This is remarkable, because for the dozens 
of GRBs further observational quantities are also known from the analyses 
of spectra and from the afterglow (AG) data. Thus, the usefulness of such 
generalized PCA is a triviality. 
This contribution presents the preliminary results of a new 
PCA, in which more variables than in \cite{ref:bagoly98} are included.

In Jochen Greiner's list \cite{ref:greiner}, 
up to September 30, 2004, there are  
38 GRBs with known redshift $z$:
for GRB980329 and GRB011030X there are upper limits
only, and in one case $z$ is queried (GRB980326). 
These 41 GRBs define the primary sample here.

For every GRB in our sample 
we will consider 8 variables: $z$, $\alpha$,
$\beta$, AG${}_{break}$, $E_{peak}$, AG$\alpha$, $T_{90}$ and fluence. 
The last two variables are the usual $T_{90}$ durations and fluences used 
already in BATSE Catalog.
The indices $\alpha$ and $\beta$ are the slopes in Band's spectrum, 
$E_{peak}$ defines the photon energy, where the Band
spectrum $EF(E)$ peaks ($F(E)dE$ is the fluence from infinitesimal photon 
energy interval $E, (E+dE)$). These three variables define the spectrum of 
GRBs. AG${}_{break}$ is the observed time (in days), when the power-law 
decay of AG
changes into a faster decay; the slope of the decay before this
AG${}_{break}$ is denoted as AG$\alpha$. (This slope is usually denoted also as
$\alpha$, but using this notation there would be a confusion with the 
spectrum's slope $\alpha$.)

All this means that we have taken only two variables from BATSE Catalog
(the $T_{90}$ duration and fluence), which in essence defined the two PCs 
in \cite{ref:bagoly98}. The third PC of BATSE Catalog, namely the fluence at 
the highest energy channel, will not be considered here, because for this 
channel the data are often vanishing and/or noisy. Our main aim 
here is to obtain informations and conclusions about the six remaining 
new quantities, not about the BATSE values themselves; hence, it is better 
not to consider here the fluence at the highest energy channel. 

There are 16 GRBs from the mentioned 41 GRBs that have all these
variables in the Greiner's list. These 16 GRBs define the net sample for our
PCA. For fluence, $T_{90}$, AG${}_{break}$ (days),
$1+z$ and $E_{peak}$ we will take the logarithms of the measured values. 

\section{Principal Component Analysis}

PCA needs, first, the calculation of correlations
among the variables. The following pairs (with the 
corresponding probabilities $p$) give strong correlations:
$\log T_{90}$ - $\log(1+z)$       ($p=66.49\%$),
$\log(1+z)$ - $\alpha$     ($p=99.75\%$),
$\log \mbox{\em Fluence}$ - $\log T_{90}$     ($p=95.49\%$),
$\log E_{peak}$ - $\log T_{90}$       ($p=97.98\%$) and 
$\log E_{peak}$ - $\log \mbox{\em Fluence}$   ($p=99.72\%$). 
All other correlations are weaker.

The second step in PCA is to calculate the eigenvalues and 
eigenvectors. PCA gives us the eigenvalues and the 
eigenvectors of the ``best'' projection (Table 1).. 

Then the third step in PCA is to define the important PCs. In accordance 
with \cite{ref:jollife} the important PCs are that ones, which have 
eigenvalues above $1.0$. From Table 1 We obtain the result that only 3 
PCs are important. This is a 
highly remarkable conclusion: we have 8 variables, and their reduction to 
3 is an essential lowering. In addition, two PCs were expected already
from the earlier paper \cite{ref:bagoly98}, because these two PCs were 
given by the BATSE data alone. In other words, the 6 new quantities - not 
used already in \cite{ref:bagoly98} - define only one further new 
important PC.

\begin{table}
\caption{PCA result.}
\label{tab:pca}
{\small
\begin{tabular}{l|rrrrrrr}
&Comp1&Comp2&Comp3&Comp4&Comp5&Comp6&Comp7\\
\hline
$\log(1+z)$&0.345&-0.359&-0.399&0.301&0.278&0.348&-0.538\\
$\alpha$&0.355&-0.394&-0.283&0.336&-0.392&-0.188&0.497\\
$\beta$&-0.342&0.016&0.394&0.798&0.226&0.128&0.098\\
$\log \mbox{AG}_{break}$&-0.225&0.473&-0.560&0.023&0.132&0.135&-0.013\\
AG$\alpha$&0.381&-0.142&0.484&-0.285&0.217&0.377&0.051\\
$\log E_{peak}$&-0.427&-0.285&-0.145&-0.173&-0.218&0.716&0.313\\
$\log T_{90}$&-0.316&-0.498&-0.154&-0.212&0.640&-0.349&0.213\\
$\log \mbox{\em Fluence}$&-0.400&-0.378&0.120&-0.055&-0.442&-0.181&-0.555\\
\hline
Eigenvalue & 4.314 & 1.761 & 1.224 & 0.464 & 0.189&0.038 &  0.009
\end{tabular}
}
\end{table}

As it is seen from Table 1, 
the first three eigenvectors (Comp1-3) are the linear combination 
of $\log E_{peak}$, $\log T_{90}$, $\log \mbox{\em Fluence}$, 
$\log(1+z)$, $\alpha$, $\log \mbox{AG}_{break}$ and AG$\alpha$.
No variable alone is dominating in these PCs. All this means 
that, beyond the two PCs coming from the BATSE Catalog, there is only one 
further important PC, and this one is a complicated linear combination
of all variables. In other words: Although the correlations among the 
variables are generally weak, these correlations are enough to ensure a 
coupling among the variables.

The $\log E_{peak}$+$\log T_{90}$+$\log \mbox{\em Fluence}$ correlation 
is well-known. This correlation in essence ensures that $\log E_{peak}$
alone is not independent. The correlation $\log(1+z)$ and $\alpha$ is also
ensures that the redshift alone is also {\it not} independent - a quite 
remarkable behavior. (As it is seen on Figure 1, the positive correlation 
between $\log(1+z)$ and $\alpha$ is obvious, because the trend - 
increasing of $z$ with increasing $\alpha$ - is well seen. 
Contrary this, no such trend is seen between $z$  
$\beta$.) Similarly, the remaining 
correlations are enough to lower the number of important variables.

\begin{figure} 
\centerline{\includegraphics[width=0.6\textwidth, angle=270]{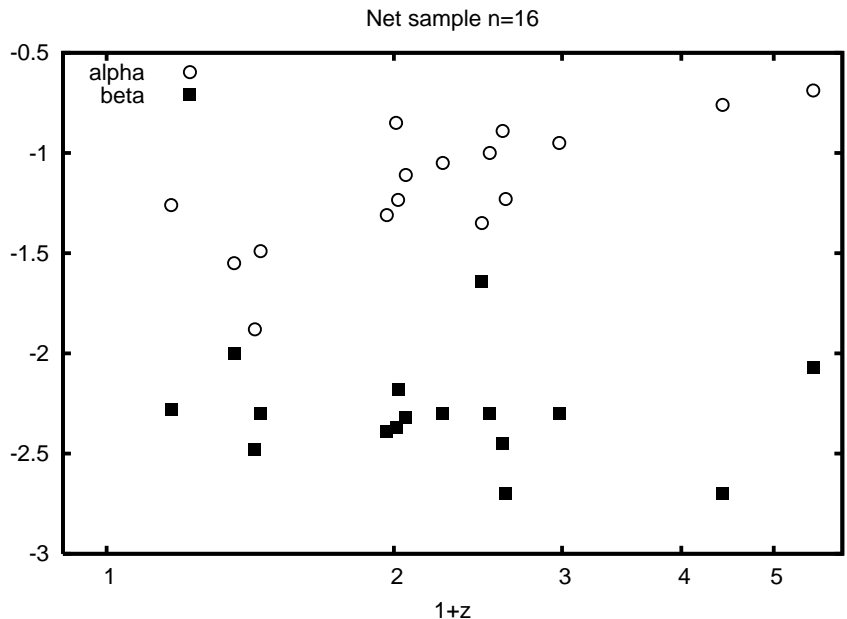}}
\caption{ $1+z$ distribution of the $\alpha$ and $\beta$ values 
of the net data.}
\end{figure} 

Note still that,
interestingly, $\beta$ stands practically alone in fourth PC 
(Comp4). Hence, the only 
variable, which can define {\it alone} a PC, is the quantity $\beta$. But, 
in accordance with \cite{ref:jollife}, this PC has already an eigenvalue 
smaller than $1.0$.

\section{Discussion}

The importance of PCA is given also by the fact that it uses exclusively 
the observational data, and any model of GRBs must respect the  
reduction of important variables. Here we briefly discuss the consequences 
of PCA on the models of GRBs.

The prompt emission in GRBs can in most cases be described by various models
of optically-thin synchrotron emission.  Theoretical considerations suggest
that, apart from this non-thermal emission, a thermal emission component, from
the photosphere of the outflow, could be equally strong in the gamma-ray
band ~\cite{ref:mesz00,ref:daigne03}. In \cite{ref:ryde04} it was found
 that the observed spectra can indeed be
described by a thermal black-body superimposed on a non-thermal spectrum. The
relative strengths of these two components vary. A large $\alpha$ is
measured if the thermal component is dominant, while a lower value (more 
like a synchrotron value) is found if the non-thermal
component dominates. The variations in $\alpha$ 
therefore could suggest a variation in importance of the photosphere.

Note that $\alpha$ values are
measured on the {\it integrated} spectra. These are necessarily 
softer than the instantaneous spectra ~\cite{ref:ryde99}.
It is the instantaneous spectrum that reflects the physical radiation 
process shaping the spectrum. We therefore have 
that $\alpha_{\rm rad. proc.} > \alpha_{\rm int.}$.
The strong correlation between $\alpha$ and $1+z$ therefore
suggests that the thermal component becomes more dominant at larger redshifts.
In \cite{ref:mesz00} it was showed that a 
combination of the variation time scale at
the central engine and the dimension-less entropy of the outflow
determines this relation. Low variability at the
central engine combined with large Lorentz factors should therefore have 
been dominant at large redshifts. \cite{ref:amati02} also found this 
$\alpha$ vs. $1+z$ correlation. 
They explained it in terms of the $E_{peak} \propto
E_{rad}^{0.5}$ correlation, where $E_{rad}$ is the total emitted energy of
GRBs.

\section{Conclusions}

PCA method indicates that three PCs are enough to describe the variability
of the data. Hence, a reduction of the eight variables into three is strongly
suggested. The strong correlation between $\alpha$ and $1+z$ suggests that the
thermal emission component becomes more dominant at larger redshifts.

\acknowledgments

Thanks are due to the valuable discussions with Claes-Ingvar Bj\"ornsson,
Istv\'an Csa\-bai, Claes Fransson, Johan Peter Uldall Fynbo, Peter 
M\'esz\'aros
and G\'abor Tusn\'ady.  This research was supported through OTKA grants T034549
and T48870, and by a grant from the Wenner-Gren Foundations (A.M.).

\end{document}